# Agile and Broadband All-Optical Wavelength Conversion with Multi-Wavelength Lasers

PABLO MARIN-PALOMO[*], SHAHAB ABDOLLAHI, MATHIEU LADOUCE, AND MARTIN VIRTE

*Brussels Photonics Team (B-PHOT), Vrije Universiteit Brussel, Pleinlaan 2, 1050 Brussel, Belgium*
*[*]pablo.marin-palomo@vub.be*

**Abstract:** We demonstrate a novel approach to all-optical wavelength conversion (AOWC) using a monolithically integrated InP multi-wavelength laser (MWL). By exploiting carrier-induced gain modulation and intermodal coupling within the common gain section, we achieve data wavelength conversion over a 1.3 THz range for signals up to 10 GBd, without the need for an external probe laser. The scheme relies on optical injection of the data signal into the MWL, where strong mode coupling enables transfer of the modulation to other longitudinal modes. A monolithically integrated feedback cavity provides agile switching between three longitudinal modes or even broadcasting across the three channels by adjusting the feedback phase. We evaluate the signal quality of the converted data through BER measurements for various symbol rates, showing both transparent conversion and, at low injection powers, even net signal gain. Complementary numerical simulations, based on a multimode extension of the Lang-Kobayashi rate equations, reveal the role of modal gain imbalance and cross-saturation in shaping the conversion efficiency. These results establish a scalable, compact, and energy-efficient route toward agile AOWC devices.

## 1. Introduction

With the continuous exponential growth of global data traffic, driven by cloud computing, AI workloads, and streaming, optical networks must scale in both capacity and flexibility [1]. Wavelength division multiplexing (WDM) has become the cornerstone of high-capacity optical communications by enabling transmission of multiple data channels in the C band through a single fiber. However, the flexibility of WDM networks is constrained by the need for wavelength continuity in traditional fixed-wavelength paths [2,3].

To enable dynamic bandwidth allocation, avoid wavelength congestion, and achieve flexible rerouting, it is essential to decouple the signal from its carrier wavelength. In this sense, wavelength conversion allows optical signals to be transferred from one wavelength to another, enabling transparent and reconfigurable optical networks for efficient spectral utilization [4,5]. Wavelength conversion, therefore, is a key enabling function for WDM networks. In conventional optical networks, routing and switching of the channels often involve optoelectronic conversion, where signals are converted from optical to electrical and back again. This process introduces significant latency, power consumption, and hardware complexity [6], limiting its suitability for the increasing demands of current high-capacity, low-latency networks. These bottlenecks become particularly acute in metro-core, long-haul, and data center networks, where hundreds or thousands of wavelength channels may need to be flexibly managed.

To overcome these limitations, all-optical wavelength conversion (AOWC) techniques have been recently widely explored, particularly on integrated platforms to reduce footprint, cost, and power consumption. These techniques mostly involve nonlinear optical processes within a semiconductor amplifier (SOA) [7,8] or a nanophotonic waveguide [9]. Among the most common approaches are four-wave mixing (FWM), difference-frequency generation (DFG) [10], cross-gain modulation (XGM), cross-phase modulation (XPM) [11], and period-one (P1) nonlinear dynamics [12]. FWM and DFG can achieve transparent, bit-rate-independent wavelength conversion and support advanced modulation formats. However, they typically require external probe lasers, stringent phase matching, and suffer from low conversion efficiencies, making them unsuitable for compact, scalable, and energy-efficient deployment. Alternatively, XGM in SOA offers a simpler, phase-matching-free configuration and can even provide signal amplification during the conversion process [13]. Nevertheless, XGM-based schemes still depend on dedicated pump or probe lasers for each conversion event, which adds complexity and limits integrability. Moreover, in these schemes, the reconfiguration speed is limited by the tuning speed of the external pump/probe laser, typically in the µs–ms range.

In this work, we propose a novel approach that exploits mode coupling within a common gain section of a custom-designed InP multi-wavelength laser (MWL) for intermodal data transfer [14–17]. Our mechanism is similar to XGM in SOAs, but instead of injecting an external pump/probe pair, here the modulation is transferred between the intrinsic modes of the MWL. We therefore refer to it as laser-assisted XGM (LA-XGM).

In our LA-XGM approach, the injected data induces a gain modulation that is transferred to the other lasing modes through intermodal coupling, i.e., the modes of the MWL themselves compete for the common carrier reservoir. Using this approach, we demonstrate wavelength conversion of an on-off keying (OOK) signal at symbol rates of up to 10 GBd from a wavelength of 1537.5 nm to a wavelength of 1548.2 nm, corresponding to approximately 1.3 THz frequency offset. In the case of low-power signals, we observe gain during the wavelength conversion process. In addition, a feedback cavity monolithically integrated with the MWL enables agile control over the wavelength conversion process, allowing for selection of the output wavelength and even data broadcasting across various wavelengths. Finally, numerical simulations based on the multi-mode extension of the Lang-Kobayashi rate equations validate the experimental results while predicting system behavior in unexplored regions. In particular, we show the role of mode coupling and mode gain imbalance on the quality of the wavelength-converted data. Overall, we believe our approach addresses fundamental limitations of conventional AOWC methods and opens new possibilities for fully integrated, reconfigurable WDM systems.

## 2. Agile all-optical wavelength conversion with an on-chip single-gain MWL

Our approach to agile all-optical wavelength conversion (AOWC) is based on optical injection of the data channel into the multi-wavelength laser (MWL), while the monolithically integrated feedback cavity enables agile control of the converted wavelength. Figure 1 shows a conceptual diagram of a WDM system with our MWL-based AOWC. The channels from the WDM signal to be wavelength-converted are demultiplexed and routed towards our feedback-controlled MWL. The feedback phase is directly controlled through a voltage signal, which allows time control of the wavelength at which the data is transferred. A wavelength-selective switch (WSS) is then used to multiplex the wavelength-converted data with the added channels in this node ($\lambda_{in}$) and those channels at a fixed wavelength. This configuration enables agile, energy-efficient, and fully integrated data wavelength conversion over wide spectral ranges (up to several THz), thus supporting flexible WDM routing and dynamic bandwidth allocation.

The core of our all-optical wavelength conversion (AOWC) is our MWL, fabricated on a generic InP photonic platform that integrates all active and passive components on a single chip [18]. The MWL is defined by a dual-cavity structure, with two distributed Bragg reflectors

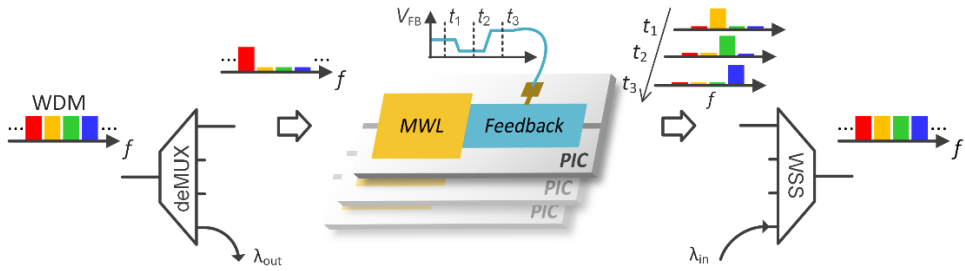

Figure 1. Conceptual representation of a WDM node employing agile all-optical wavelength conversion with an on-chip feedback-controlled multi-wavelength laser (MWL). The channels from the WDM signal to be wavelength-converted are demultiplexed and routed towards our feedback-controlled MWL. The feedback phase is directly controlled through a voltage signal, which allows time control of the wavelength at which the data is transferred. A wavelength-selective switch (WSS) is then used to multiplex the wavelength-converted data with the added channels in this node ($\lambda_{in}$) and those channels at fixed wavelength

(DBR) on one side and a broadband 50-50 two-port multimode interference reflector (MIR) on the other side, as illustrated in Fig. 2(a). See Fig. 2(b) for the image of the photonic integrated circuit (PIC) and of the electrically packaged PIC including the lensed fiber (LF) which is used to couple the light in and out of the PIC. The two DBR structures are placed in series, with DBR1 defining the shortest cavity with a length of 1082.43 µm, and DBR2 the longest cavity with a length of 1396.3 µm. The pitch is set to 0.236 µm for DBR1 and 0.234 µm for DBR2, this corresponds to Bragg wavelengths detuned from each other by approximately 10 nm (1.3 THz). The gain section is a 500 µm-long quantum well semiconductor optical amplifier ($SOA_1$), shared across both cavities. The laser is driven at 50 mA (approximately twice the lasing threshold) and operated at a temperature of 22 °C unless otherwise indicated. In this configuration, the laser supports multiple longitudinal modes, see Fig. 2(c). The dominant modes, centered at $\lambda_1$ = 1547.94 nm, $\lambda_2$ = 1548.21 nm, and $\lambda_3$ = 1548.49 nm, arise from the cavity defined by DBR1 and are spaced by approximately 30 GHz. The suppressed modes, i.e., longitudinal modes whose power levels are at least 20 dB lower than the dominant modes at any operation condition, are centered at around 1537 nm and are associated with the cavity defined by DBR2. Note that the modal gain difference between the modes associated with DBR1 and DBR2 can, to some extent, be adjusted by injecting current into the DBRs [16]. However, during our experiments, no current is applied to the DBRs.

The MWL is coupled to an external feedback cavity through an 85/15 multimode interference (MMI) splitter, such that 15% of the output light is directed toward the feedback loop. This external cavity comprises a 300 µm-long $SOA_2$, a 1200-µm long electro-optic phase modulator (EOPM) with $V_\pi$ = 8 V, and a one-port MIR with approximately 80% reflectivity. The length of the feedback loop is approximately 2420 µm long, which largely places us in the so-called short-cavity regime [19]. This feedback control enables nanosecond-scale switching between converted wavelengths [16].

Both the MWL and the optical feedback (OFB) circuitry are monolithically integrated on a single InP photonic chip [14], which is electrically packaged with all contact pads wire-bonded to PCB boards, see Fig. 2(b). The chip is mounted on a Peltier element equipped with a thermistor for precise temperature control.

The emission of the MWL when the $SOA_1$ is operated at 50 mA consists of multiple modes, see Fig. 2(c). Applying a relatively low current of 9 mA to $SOA_2$ in the external cavity increases the power of the feedback signal enough to influence the mode competition and enables control of the MWL emission, i.e., it allows selecting which of the modes dominates the emission. The control relies on adjusting the phase of the feedback signal with a voltage-controlled EOPM. Through the external Fabry-Pérot cavity, each mode emitted by the MWL accumulates a round-trip phase proportional to the voltage applied to the EOPM. In this configuration, the in-phase modes experience a gain boost as opposed to the out-of-phase modes, which will comparatively experience higher losses [14].

Figure 2(d) depicts the power of each of the three dominant modes as a function of the applied voltage to the EOPM, $V_{EOPM}$. For $V_{EOPM}$ values below 2 V, the MWL emission is dominated by the mode at $\lambda_2$, between approximately 2 V and 3.5 V, $\lambda_3$ dominates, while above 4 V, $\lambda_1$ dominates. Between approximately 3.5 V and 4 V, the three main dominating modes are emitted with relatively equal power levels. For $V_{EOPM}$ values above $V_\pi$, the phase accumulated through the feedback cavity is $2\pi$, and $\lambda_2$ dominates once more.

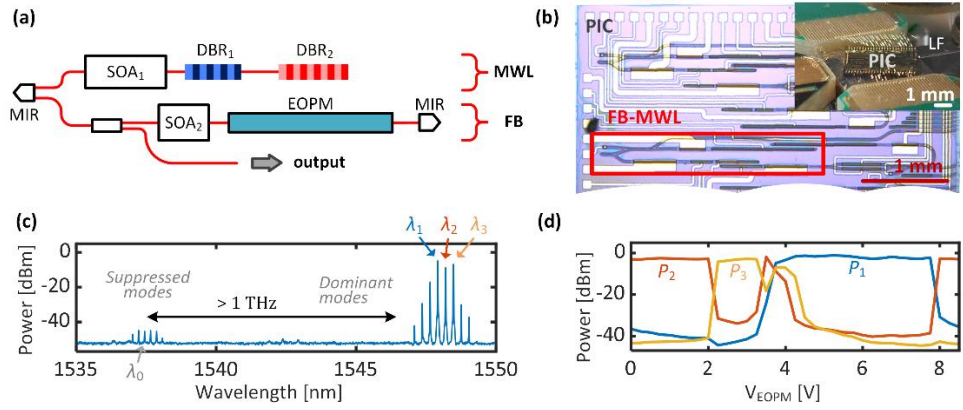

Fig. 2. Feedback-based emission control from an on-chip multi-wavelength laser (MWL). (a) Scheme describing the components of our MWL. DBR: Distributed Bragg reflector; SOA: Semiconductor optical amplifier; MIR: Multimode interference reflector; EOPM: Electro-optic phase modulator; LF: Lensed fiber; OSA: Optical spectrum analyzer. (b) Image of the PIC containing several MWL structures. (c) Optical spectrum of MWL output for IFB = 9 mA and VEOPM = 3.75 V. (d) MWL output power as a function of the FB voltage $V_{EOPM}$ for $I_{FB}$ = 9 mA, $\lambda_1$ = 1547.94 nm, $\lambda_2$ = 1548.21 nm, and $\lambda_3$ = 1548.49 nm.

## 3. All-optical wavelength conversion over 1.3 THz of OOK signals

All-optical wavelength conversion of the data injected into our MWL is achieved via laser-assisted cross-gain modulation (LA-XGM), involving carrier-induced gain modulation and intermodal coupling, but without the need for external probe lasers. The data channel at the wavelength to be converted is optically injected into the cavity of the MWL, which is set such that one of its suppressed modes, see Fig. 2(c), overlaps with the injected signal. The optical injection depletes the carriers due to stimulated emission. In case of a modulated signal this leads to a temporal modulation of the carrier density, which in turn modulates the gain experienced by other modes of the MWL. In our MWL, the fact that all longitudinal modes share a single gain section leads to the intensity modulation of the un-injected modes, thereby transferring the data onto the spectrally distinct lasing modes. This mechanism allows us to wavelength-convert data signals to any wavelength emitted by the MWL, i.e., with a spectral separation up to several THz, without the need for any additional pump or probe lasers, since our laser structure naturally supports multi-mode operation. Importantly, injecting the optical signal at a suppressed mode of the MWL and wavelength-converting it to a dominant mode increases the conversion efficiency.

The experimental setup for AOWC of data signals is depicted in Fig. 3(a). The electrical OOK signal is generated using an arbitrary waveform generator (AWG, Keysight M8194A). We consider a pseudo-random bit sequence of length $2^{11}$-1 and use a root-raised cosine pulse shaping with a 10% roll-off. A Mach-Zehnder modulator (MZM) encodes the signal onto an optical carrier at $\lambda_0$ = 1537.5 nm, obtained from a high-quality tunable laser (Keysight N7776C). The optical power and polarization of the signal are adjusted using a variable optical attenuator (VOA) and a polarization controller (PC), respectively. The signal is coupled into and out of the chip using a lensed fiber (LF) with approximately 3 dB of coupling losses. An optical fiber circulator (OFC) is used to separate the injected signal from the output signal from the MWL. We measure the output optical spectrum with an optical spectrum analyzer (OSA). To retrieve the wavelength-converted data, we first send the output signal through a bandpass filter (BPF) with a 3-dB bandwidth of approximately 30 GHz, which filters out the desired channel. Next, the filtered signal is sent to a photoreceiver (PR, Thorlabs RXM42AF)

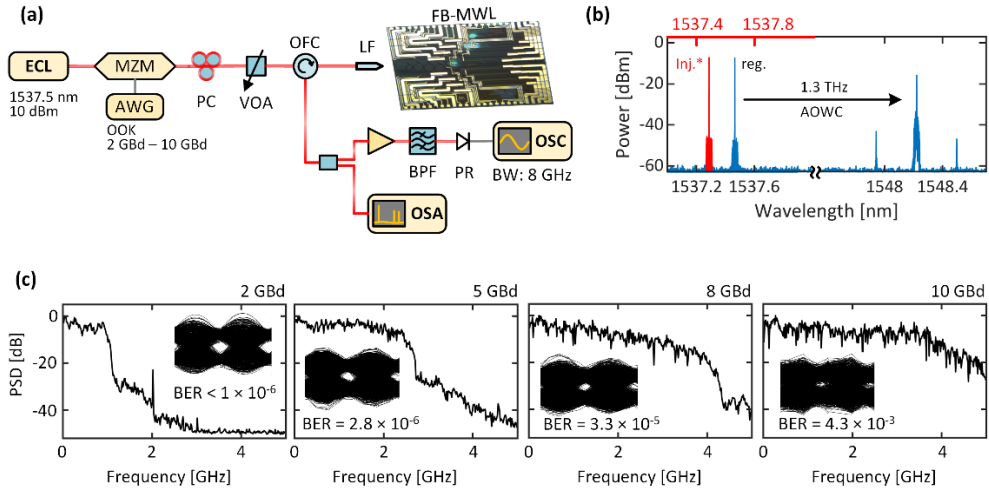

Fig. 3. Data wavelength conversion in an MWL. (a) Setup for data wavelength conversion. MZM: Mach-Zender modulator; AWG: Arbitrary waveform generator; PC: polarization controller; VOA: Variable optical attenuator; OFC: Optical fiber circulator. LF: Lensed fiber. BPF: Band-pass filter. PD: Photodetector. (b) Exemplary optical spectrum of the light emitted by the MWL when a 5GBd OOK signal is optically injected around 1537.5 nm. The wavelength-converted signal is observed at $\lambda_2$ = 1548.21 nm. (c) Power spectrum of the OOK signal recorded by the oscilloscope for different symbol rates

connected to an oscilloscope (Teledyne Lecroy WavePro HD). An erbium-doped fiber amplifier (EDFA) set to an output power of 5 dBm is employed to compensate for the coupling losses and the insertion loss from the BPF. The signal from the oscilloscope is processed to extract the BER from the eye diagram [20].

To demonstrate AOWC, we adjust the feedback phase so that the data is wavelength-converted from $\lambda_0$ = 1537.5 nm to $\lambda_2$ = 1548.21 nm, and we analyze the impact of the symbol rate on the signal quality. We investigate symbol rates ranging from 2 GBd to 10 GBd. Figure 3(b) depicts exemplary optical spectra of the injected (inj.) 5 GBd OOK signal with an on-chip injected optical power of approximately – 7 dBm (red), and of the output signal from the MWL (blue). The frequency axes of the spectra are artificially offset by 0.2 nm to facilitate the visualization. The spectrum of the output signals contains both the regenerated (reg.) signal associated with the amplification of the injected signal and the wavelength-converted OOK signal 10 nm (1.3 THz) shifted towards longer wavelengths, at $\lambda_2$. In this configuration, the converted channel exhibits a larger modulation depth and higher output power. While in Section 4, we investigate the impact of the injection power, a systematic characterization of the resulting modulation depth is yet to be performed

Figure 3(c) depicts the power spectral density (PSD) of the time-domain signal directly extracted from the oscilloscope for various symbol rates together with the corresponding eye diagram and BER. A clear modulation envelope and eye-opening is observed, indicating successful wavelength conversion. Although the bandwidth of the converted signal degrades at higher data rates, we are able to recover signals up to 10 GBd. In our experiments, the limited performance at higher symbol rates is attributed to the finite XGM bandwidth, which is fundamentally constrained by carrier recovery times or the effective carrier lifetime of a few hundred ps in InP QWs [21], corresponding to a few GHz, and the relaxation oscillation frequency of the lasing modes [22,23], which is only a few GHz in our devices. In [22], the authors show that optimizing the length of the gain section and the gain saturation, bandwidths of up to 20 GHz could be achieved. Furthermore, optical injection locking or near-locking operation has been shown to significantly increase damping of relaxation oscillations and shorten the effective carrier lifetime, thereby extending modulation bandwidth to tens of GHz both in quantum-dot and quantum well devices [24,25].

In summary, with our approach, signal quality, bandwidth limitations, and conversion efficiency are sensitive to both optical injection and MWL parameters. Injection parameters include the injection strength and detuning, i.e., the wavelength difference between the injected signal and that of the suppressed mode of the MWL. Regarding the MWL, key parameters are, among others, the injection current, modal gains, and modal coupling.

At this stage, although we have not performed a systematic experimental analysis of the impact of these parameters on the signal quality and the achievable bandwidth, numerical investigations, see Section 5, already identify operating conditions that maximize conversion bandwidth and minimize distortion.

## 4. Dynamic all-optical wavelength conversion

One of the key advantages of our integrated MWL-based approach is the ability to dynamically control the wavelength to which data is converted. To achieve dynamic control over the data output wavelength, we optically inject the OOK signal near one of the suppressed longitudinal modes of the MWL, as described in Section 3. Due to strong mode coupling within the common gain medium, this injected modulation is replicated onto the dominant FP modes of the MWL, separated by more than 1.3 THz from the injection wavelength. In our case, data initially injected at $\lambda_0$ = 1537.5 nm is transferred to $\lambda_1$ = 1547.94 nm, $\lambda_2$ = 1548.22 nm, and $\lambda_3$ = 1548.50 nm, depending on the feedback conditions. Next, mode selection is obtained by adjusting the phase of the optical feedback, which is implemented via the monolithically integrated feedback cavity described in Section 2. The EOPM allows precise tuning of the feedback phase by varying the applied voltage, while the SOA in the feedback loop adjusts the feedback strength. Together, these components control the modal gain competition, enabling reconfigurable selection of the output channel's wavelength.

We demonstrate dynamic control with a 5 GBd OOK signal. Figure 4(a) depicts the spectra of the output from the MWL, including the regenerated and the wavelength-converted signals for different $V_{\text{EOPM}}$ values. Figure 4(b) shows the optical power associated with each of the channels. We observe distinct switching events, marked by abrupt changes in the dominant emission mode and corresponding BER performance, see Fig. 4(c). The measurements reveal that data recovery can be achieved in either mode, depending on the chosen feedback phase. Importantly, we have observed that feedback-based mode switching can occur within the nanosecond regime [16,26], which is essential for high-speed reconfigurability.

We further investigate the impact of system parameters on the signal quality. Figure 4(d) illustrates the measured BER as a function of injection strength and feedback SOA current. Although the analysis does not cover a broad parameters space, these preliminary results suggest that lower injection strengths (i.e., weaker saturation of the gain medium by the input signal) and higher feedback strength (via increased SOA current) tend to improve signal quality, likely due to more favorable gain redistribution dynamics. These results motivate a more comprehensive parametric study to optimize the system for higher data rates and lower BER.

A particularly interesting feature of our platform is its ability to support simultaneous data broadcasting across multiple wavelengths. For certain feedback phase settings, the system enters a regime where three FP modes lase simultaneously, with each mode carrying the same modulated data signal. This wavelength-multiplexed broadcasting effect is illustrated in Fig. 4(e). Such behavior could be harnessed in applications requiring optical fan-out, data duplication, or multi-channel replication in WDM-based networks [27].

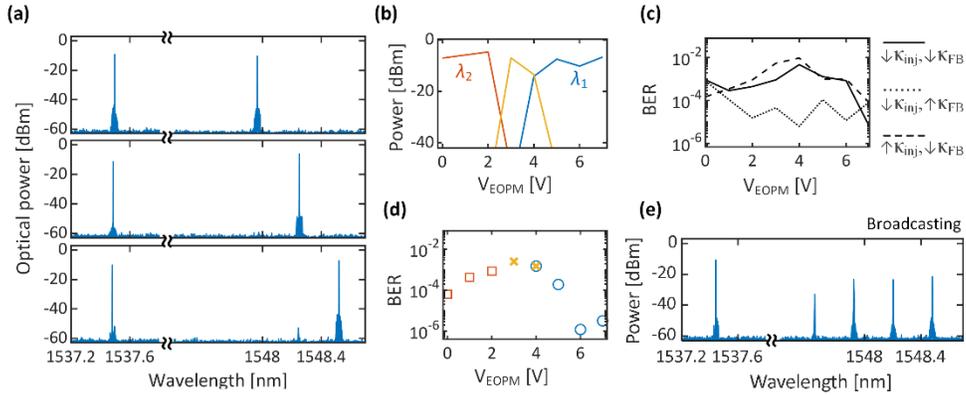

Fig. 4. Dynamic all-optical wavelength conversion of a 5 GBd OOK signal at $\lambda_0$ = 1537.5 nm with – 7 dBm on-chip power. (a) Optical spectrum of the MWL output when applying 21 mA to the $SOA_2$ in the FB cavity and varying the $V_{EOPM}$. (b) Optical power and (c) BER associated with each of the channels. (d) Impact of feedback strength and injection strength on the signal quality of the channel with the highest optical power at each $V_{EOPM}$. (e) Spectrum of the output of the MWL for – 4 dBm on-chip injection power and applying 9 mA to $SOA_2$. The three modes carry the same OOK data.

## 5. Modeling and Numerical Simulations

To validate our experimental results and explore system behavior beyond the range of experimentally accessible parameter values, we perform numerical simulations of the MWL dynamics under optical injection and feedback. These simulations are based on a multi-mode expansion of the Lang-Kobayashi equations [28], modified to include modulated optical injection, following the formalism described in [16], to represent the injected data signal. A detailed description of the model and its parameters is provided in the Methods section. In this section, we investigate how various laser and injection parameters affect the performance of data wavelength conversion.

The simulations consider the injection of an OOK-modulated signal with a symbol rate of 5 GBd into mode 1 and evaluate the BER of the converted signal at mode 2. Throughout the simulations, we fix the following parameters: $\alpha$ = 3 (linewidth enhancement factor), $\kappa$ = 0.08 (feedback strength), T = 1000 (carrier-to-photon lifetime ratio), and P = 3 (normalized pump level). Figure 5(a) presents the resulting BER maps as a function of the frequency detuning $\Delta$ and the injection strength $\kappa_{inj}$, for different combinations of modal gain difference $g_2 - g_1$ and cross-saturation parameter $\beta$.

The results reveal key trade-offs in the injection conditions: At low injection strength values, the weak modulation depth transferred to the un-injected mode results in poor signal quality. At high injection power values, the laser tends to dynamically lock to the injected mode, reducing power at the target (converted) mode, and thereby degrading conversion efficiency. For high injection strength values, we also observe the onset of dynamics leading to a distorted wavelength-converted signal. These regions are highlighted as blank regions in Fig. 5(a). These findings are consistent with our experimental observations in Sections 3 and 4, where the excessive injection strength leads to either dynamic instabilities or full injection locking, suppressing the converted mode, while too weak injection results in insufficient power transfer. Regarding the detuning, optimal performance occurs within a moderate region asymmetrically distributed around zero detuning due to the influence of the non-zero α-factor. Figure 5(b) showcases the power spectrum around the un-injected mode for constant detuning and increasing injection strength. Regarding the laser parameters, we observe that low modal gain difference values enable the use of lower injection strength values. In addition, we observe that the coupling value largely impacts the shape of the region where wavelength conversion can be achieved.

Although it is beyond the scope of this paper, our findings suggest that careful optimization of the laser design, particularly the cavity geometry, gain asymmetry, and intermodal coupling, can substantially expand the performance envelope of our wavelength conversion system. While our current experimental implementation operates under specific design constraints, the simulations indicate a broad and mostly unexplored parameter space with potential for improved signal quality and higher symbol rates.

In summary, our numerical simulations not only reinforce the experimental observations but also highlight key directions for enhancing the capabilities of integrated all-optical wavelength converters through design-aware optimization and broader parameter exploration.

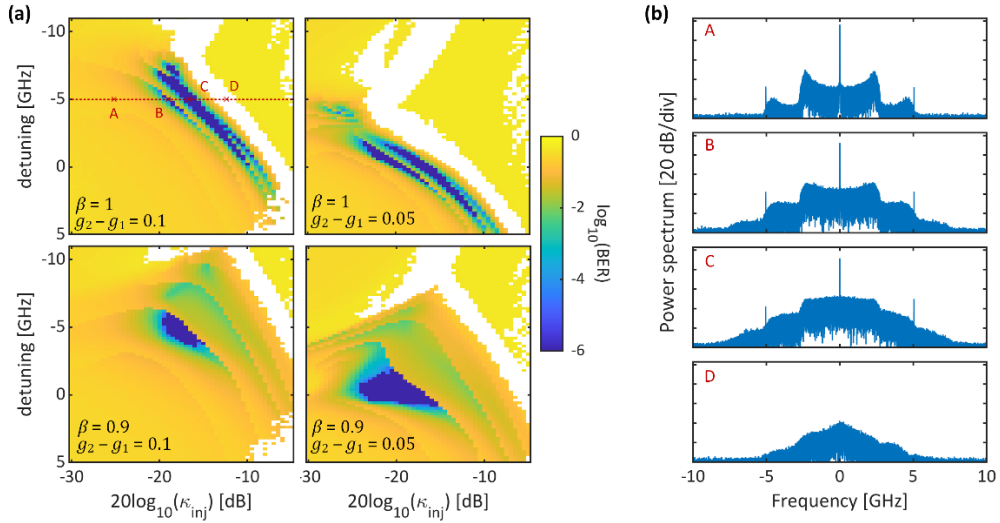

Fig. 5. Numerical investigation of the quality of data wavelength conversion of a 5 GBd OOK signal. (a) BER as a function of the detuning and injection strength for different values of the cross-saturation parameters and the modal gain difference. (b) Exemplary power spectrum of the wavelength-converted data for $\beta = 1$, $g_1 = 0.9$, $g_2 = 1$, and different $\kappa_{inj}$ values.

## 6. Conclusion

We have demonstrated all-optical wavelength conversion (AOWC) of data signals using a monolithically integrated multi-wavelength laser (MWL) fabricated on an InP platform. Our approach leverages cross-gain modulation (XGM) within a single gain section to transfer modulated data from an injected signal to other longitudinal modes of the MWL, achieving wavelength conversion over a spectral offset of 1.3 THz. Using a monolithically integrated feedback cavity, composed of a semiconductor optical amplifier (SOA) and an electro-optic phase modulator (EOPM), we enable fast and tunable wavelength selection, with switching times on the nanosecond scale.

We demonstrated conversion of OOK data up to 10 GBd, with the ability to dynamically select or even broadcast data across three output wavelengths. Although this value is low compared with FWM approaches for AOWC, achieving up to 40 GBd [9], it is comparable with conventional XGM approaches, with up to 10 GBd [29], to the best of our knowledge. In addition, through numerical simulations based on a multimode rate-equation model, we analyzed how modal gain imbalance and cross-saturation dynamics affect the signal quality and conversion efficiency.

While our approach offers significant advantages in integration, tunability, and performance, some limitations remain. The conversion bandwidth is currently constrained by modal dynamics and gain saturation, requiring further optimization of both the injection conditions and the laser cavity design. Nevertheless, our numerical investigation suggests that with optimized injection conditions, our MWL platform could credibly support wavelength conversion of higher-baud-rate signals, potentially reaching tens of GHz, provided the injection conditions and laser design are further optimized.

Future extensions of this work will investigate the feasibility of XPM-based conversion for coherent modulation formats, and the design of laser architectures supporting more modes and broader spectral spans. These developments will help establish a path toward scalable, reconfigurable, and energy-efficient all-optical wavelength converters, suited for next-generation WDM networks and data center interconnects.

Given the scalable nature of our MWL approach, we believe that by optimizing the laser cavity design, it is possible to extend the system to support a larger number of modes and higher symbol rates over even larger frequency offsets, ultimately limited by the SOA gain bandwidth, which exceeds 5 THz in InP-based platforms [18]. By enabling on-chip, dynamically tunable, and scalable all-optical wavelength conversion, our approach addresses fundamental limitations of conventional AOWC methods and opens new possibilities for fully integrated, reconfigurable WDM systems

**Funding.** This work was supported by the Research Foundation Flanders (FWO grant 1275924N, PMP) and the European Research Council (ERC, Starting Grant COLOR'UP 948129, MV).

**Data availability.** Data underlying the results presented in this paper are available in xxx.

## Methods

### A. Modeling of the MWL under feedback and modulated optical injection

We use the multimode rate equation model introduced by [30,31], which is a multi-mode extension of the Lang-Kobayashi rate equations [32], already normalized in time. We add an extra term in the field equation to account for modulated optical injection [16]. To simplify the numerical treatment, we transform the equations into an autonomous system of delay-differential equations (DDEs) by shifting the reference frame to the carrier frequency of the injected signal. This eliminates explicit time dependence in the modulation term, allowing efficient exploration of steady-state and transient dynamics. These equations read as:

$$\frac{dF_1}{dt} = (1 + i\alpha)\left(g_1 N_1 - \frac{1-g_1}{2}\right) F_1 + \kappa m(t) e^{i\Delta t}, \quad (1)$$

$$\frac{dE_2}{dt} = (1 + i\alpha)\left(g_2 N_2 - \frac{1-g_2}{2}\right) E_2, \quad (2)$$

$$T \frac{dN_1}{dt} = P - N_1 - (1 + 2N_1)(g_1 |E_1|^2 + g_2 \beta |E_2|^2), \quad (3)$$

and

$$T\frac{dN_2}{dt} = P - N_2 - (1 + 2N_2)(g_1\beta|E_1|^2 + g_2|E_2|^2). \quad (4)$$

Where $F=Ee^{-i\Delta t}$, with $E$ being the normalized electric field and $\Delta$ the normalized injection detuning between the center frequency of the injection signal and that of the injected mode. $N$ is the normalized carrier population, $\alpha$ is the linewidth enhancement factor, $g$ is the modal gain, $\kappa$ is the normalized injection strength, $\beta$ is the cross-gain saturation, $P=(J-J_{th})/2J_{th}$ is the normalized injection current, with $J_{th}$ being the threshold current, $T=\tau_c/\tau_p$, with $\tau_c$ being the carrier lifetime and $\tau_p$ the photon lifetime, and $m(t)$ is the OOK injection signal.

**References**

1.  I. Redpath, "Toward 100 Tbps and a Simplified All-Optical Network," (2024).
2.  R. Ramaswami and K. N. Sivarajan, "Routing and wavelength assignment in all-optical networks," IEEE/ACM Transactions on Networking **3**, 489–500 (1995).
3.  H. Zang, J. P. Jue, and B. Mukherjee, "A Review of Routing and Wavelength Assignment Approaches for WavelengthRouted Optical WDM Networks," Opt. Networks Mag **1**, (2000).
4.  K. Ishii, T. Inoue, I. Kim, X. Wang, H. N. Tan, Q. Zhang, T. Ikeuchi, and S. Namiki, "Analysis and Demonstration of Network Utilization Improvement Through Format-Agnostic Multi-Channel Wavelength Converters," J. Opt. Commun. Netw., JOCN **10**, A165–A174 (2018).
5.  E. Ciaramella, "Wavelength Conversion and All-Optical Regeneration: Achievements and Open Issues," J. Lightwave Technol., JLT **30**, 572–582 (2012).
6.  L. Eldada, "Advances in ROADM technologies and subsystems," in *Photonic Applications in Devices and Communication Systems* (SPIE, 2005), Vol. 5970, pp. 611–620.
7.  A. Sobhanan, A. Anthur, S. O'Duill, M. Pelusi, S. Namiki, L. Barry, D. Venkitesh, and G. P. Agrawal, "Semiconductor optical amplifiers: recent advances and applications," Adv. Opt. Photon. **14**, 571 (2022).
8.  S. J. B. Yoo, "Wavelength conversion technologies for WDM network applications," Journal of Lightwave Technology **14**, 955–966 (1996).
9.  P. Zhao, Z. He, V. Shekhawat, M. Karlsson, and P. A. Andrekson, "100-Gbps per-channel all-optical wavelength conversion without pre-amplifiers based on an integrated nanophotonic platform," Nanophotonics **12**, 3427–3434 (2023).
10. A. Schlager, M. Götsch, R. J. Chapman, S. Frick, H. Thiel, H. Suchomel, M. Kamp, S. Höfling, C. Schneider, and G. Weihs, "Difference-frequency generation in an AlGaAs Bragg-reflection waveguide using an on-chip electrically-pumped quantum dot laser," J. Opt. **23**, 085802 (2021).
11. M. S. Ab-Rahman and A. A. Swedan, "Quadruple multi-wavelength conversion for access network scalability based on cross-phase modulation in an SOA-MZI," Open Physics **15**, 1077–1085 (2017).
12. S.-K. Hwang, H.-F. Chen, and C.-Y. Lin, "All-optical frequency conversion using nonlinear dynamics of semiconductor lasers," Opt. Lett., OL **34**, 812–814 (2009).
13. K. Obermann, S. Kindt, and D. Breuer, "Performance Analysis of Wavelength Converters Based on Cross-Gain Modulation in Semiconductor-Optical Amplifiers," J. Lightwave Technol., JLT **16**, 78 (1998).
14. R. Pawlus, S. Breuer, and M. Virte, "Control of dual-wavelength laser emission via optical feedback phase tuning," Opt. Continuum **2**, 2186 (2023).
15. P. Marin-Palomo and M. Virte, "Over 1.3 THz Tunable All-Optical Wavelength Conversion with a Feedback-Controlled Multi-Wavelength Laser," in *2025 Conference on Lasers and Electro-Optics Europe & European Quantum Electronics Conference (CLEO/Europe-EQEC)* (2025), pp. 1–1.
16. S. Abdollahi, M. Ladouce, P. Marin-Palomo, and M. Virte, "Agile THz-range spectral multiplication of frequency combs using a multi-wavelength laser," Nat Commun **15**, 1305 (2024).
17. P. Marin-Palomo, D. Dadhich, S. Abdollahi, M. Ladouce, and M. Virte, "Data wavelength conversion over THz range using an optical-feedback-controlled InP multimode laser," in *Integrated Photonics Platforms III* (SPIE, 2024), Vol. 13012, pp. 73–77.
18. L. M. Augustin, R. Santos, E. den Haan, S. Kleijn, P. J. A. Thijs, S. Latkowski, D. Zhao, W. Yao, J. Bolk, H. Ambrosius, S. Mingaleev, A. Richter, A. Bakker, and T. Korthorst, "InP-Based Generic Foundry Platform for Photonic Integrated Circuits," IEEE Journal of Selected Topics in Quantum Electronics **24**, 1–10 (2018).
19. T. Heil, I. Fischer, W. Elsäßer, and A. Gavrielides, "Dynamics of Semiconductor Lasers Subject to Delayed Optical Feedback: The Short Cavity Regime," Phys. Rev. Lett. **87**, (2001).
20. W. Freude, R. Schmogrow, B. Nebendahl, M. Winter, A. Josten, D. Hillerkuss, S. Koenig, J. Meyer, M. Dreschmann, M. Huebner, C. Koos, J. Becker, and J. Leuthold, "Quality metrics for optical signals: Eye diagram, Q-factor, OSNR, EVM and BER," in *2012 14th International Conference on Transparent Optical Networks (ICTON)* (2012), pp. 1–4.

21. C. Qin, X. Huang, and X. Zhang, "Theoretical investigation on gain recovery dynamics in step quantum well semiconductor optical amplifiers," J. Opt. Soc. Am. B **29**, 607 (2012).
22. T. Durhuus, B. Mikkelsen, C. Joergensen, S. Lykke Danielsen, and K. E. Stubkjaer, "All-optical wavelength conversion by semiconductor optical amplifiers," Journal of Lightwave Technology **14**, 942–954 (1996).
23. K. Obermann, S. Kindt, D. Breuer, and K. Petermann, "Performance analysis of wavelength converters based on cross-gain modulation in semiconductor-optical amplifiers," J. Lightwave Technol. **16**, 78–85 (1998).
24. Z. Liu and R. Slavik, "Optical Injection Locking: From Principle to Applications," J. Lightwave Technol. **38**, 43–59 (2020).
25. X. Jin and S.-L. Chuang, "Bandwidth enhancement of Fabry-Perot quantum-well lasers by injection-locking," Solid-State Electronics **50**, 1141–1149 (2006).
26. M. Ladouce, S. Abdollahi, P. Marin-Palomo, and M. Virte, "On-chip multiwavelength lasers enabling nanosecond scale wavelength switching times," in *Integrated Photonics Platforms III* (SPIE, 2024), Vol. 13012, pp. 81–84.
27. Y. Lin, A. P. Anthur, S. P. Ó Dúill, F. Liu, Y. Yu, and L. P. Barry, "Fast Reconfigurable SOA-Based Wavelength Conversion of Advanced Modulation Format Data," Applied Sciences **7**, 1033 (2017).
28. I. V. Koryukin and P. Mandel, "Dynamics of semiconductor lasers with optical feedback: Comparison of multimode models in the low-frequency fluctuation regime," Phys. Rev. A **70**, 053819 (2004).
29. F. Bontempi, S. Faralli, N. Andriolli, and G. Contestabile, "An InP Monolithically Integrated Unicast and Multicast Wavelength Converter," IEEE Photonics Technology Letters **25**, 2178–2181 (2013).
30. E. A. Viktorov and P. Mandel, "Low Frequency Fluctuations in a Multimode Semiconductor Laser with Optical Feedback," Phys. Rev. Lett. **85**, 3157–3160 (2000).
31. I. V. Koryukin and P. Mandel, "Antiphase dynamics of selectively coupled multimode semiconductor lasers," IEEE Journal of Quantum Electronics **39**, 1521–1525 (2003).
32. R. Lang and K. Kobayashi, "External optical feedback effects on semiconductor injection laser properties," IEEE Journal of Quantum Electronics **16**, 347–355 (1980).